\documentclass[a4paper,fleqn,usenatbib]{mnras}

\usepackage{newtxtext,newtxmath}
\usepackage{ragged2e}    

\usepackage[T1]{fontenc}
\usepackage{ae,aecompl}

\usepackage{graphicx}	
\usepackage{amsmath}	
\usepackage{amssymb}	
\usepackage{natbib}
\usepackage{longtable}
\usepackage{color}

\newcommand{\asec}{$^{\prime\prime}$}

\def\H{N$_{2}$H$^{+}$}

\def\15N{$^{15}$NNH$^+$}
\def\N15{N$^{15}$NH$^+$}

\def\METH{CH$_3$OH}

\def\kms{\mbox{km~s$^{-1}$}}
\def\cmc{cm$^{-3}$}
\def\cmq{cm$^{-2}$}

\usepackage{topcoman} 
\usepackage{booktabs} 

\title[On the origin of PN in star-forming regions]{On the origin of phosphorus nitride in star-forming regions}

\author[Mininni et al.]{
C. Mininni$^{1}$\thanks{E-mail: mininni@arcetri.astro.it},
F. Fontani$^{2}$,
V. M. Rivilla$^{2}$,
M. T. Beltr\'an$^{2}$,
P. Caselli$^{3}$,
and A. Vasyunin$^{3,4}$
\\
$^{1}$Dipartimento di Fisica e Astronomia, Universit\`a degli Studi di Firenze, I-50125 Firenze, Italy \\
$^{2}$INAF/Osservatorio Astrofisico di Arcetri, Largo Enrico Fermi 5, I-50125, Florence, Italy\\
$^{3}$Centre for Astrochemical Studies, Max-Planck-Institute for Extraterrestrial Physics, Giessenbachstrasse 1, 85748 Garching, Germany  \\
$^{4}$Ural Federal University, Ekaterinburg, Russia \\
}

\date{Accepted XXX. Received YYY; in original form ZZZ}

\pubyear{2017}

\begin{document}
\label{firstpage}
\pagerange{\pageref{firstpage}--\pageref{lastpage}}
\maketitle

\begin{abstract}
We present multi-transition observations of PN towards a sample of nine massive dense cores in 
different evolutionary stages. Using transitions with different excitation conditions, 
we have found for the first time that the excitation temperatures of PN are in the range $\sim $5--30 K. 
To investigate the main chemical route for the PN formation (surface-chemistry vs. gas-phase chemistry), 
and the dominant desorption mechanism (thermal vs. shock), 
we have compared our results with those obtained from molecules tracing different chemical and 
physical conditions (SiO, SO, CH$_3$OH, and N$_2$H$^+$). We have found that the PN line profiles
are very well correlated with those of SiO and SO in six out of the nine targets,
which indicates that PN may be released by sputtering of dust grains due to shocks. 
This finding is corroborated by a faint but statistically significant positive trend between the
PN abundance and those of SiO and SO.
However, in three objects the PN lines have no hints of high velocity wings, which indicates
an alternative origin of PN. 
Overall, our results indicate that the origin of PN is not unique, as it can be formed
in protostellar shocks, but also in colder and more quiescent gas through alternative pathways.
\end{abstract}




\section{Introduction}
\label{intro}

Phosphorus (P) is one of the most important element for (pre)biotic chemistry, because it is crucial to the formation of nucleic acids, cellular membranes, and adenosine triphosphate (ATP), the key molecule for the energy transfers in cells (see e.g. Pasek \& Lauretta~\citeyear{pel}, Pasek et 
al.~\citeyear{pasek2017}). Phosphorus-bearing molecules are found in pristine solar system material such as 
asteroids (Mac\'ia~\citeyear{macia}) and comets (Altwegg et al.~\citeyear{altwegg}), whose impact (and release of their content) on the primordial Earth could have had a key role in the emergence of life as we know it. However, the interstellar chemistry of P is far less understood than that of other elements. Phosphorus is thought to be synthesized in massive stars and injected into the interstellar medium via supernova explosions (Koo et al.~\citeyear{koo}, Roederer et al.~\citeyear{roederer}). It has a relatively low cosmic abundance (P/H$\sim$ 2.8$\times$10$^{-7}$; Grevesse \& Sauval~\citeyear{ges}), and it is thought to be depleted in the dense and cold interstellar medium by a factor of 600 (e.g., Wakelam \& Herbst 2008). Because P is essentially undepleted in diffuse clouds (Lebouteiller et al.~\citeyear{lebouteiller}), depletion of P should be due to freeze-out onto the icy mantles of dust grains, and its desorption mechanisms should be similar to those of all the other icy mantle components. Among the phosphorus-bearing molecules, the phosphorus nitride (PN) is the first one detected in the interstellar medium toward three high-mass star-forming regions: Orion KL, Sgr B2, and W51, in which the measured abundances are $\sim$ (1--4) $\times$ 10$^{-10}$, larger than theoretically expected from a pure low-temperature ion-molecule chemical network (Turner \& Bally~\citeyear{teb}, Ziurys~\citeyear{ziurys}). Since then, it has been detected in high-mass dense cores (Turner et al.~\citeyear{turner}, Fontani et al.~\citeyear{fontani2016}), as well as in the circumstellar material of carbon- and oxygen-rich stars (e.g., Milam et al.~\citeyear{milam}, De Beck et al.~\citeyear{debeck}) and in protostellar shocks (Lefloch et al.~\citeyear{lefloch}). Other phosphorus-bearing molecules (e.g., CP, HCP, PH$_3$) have been detected in evolved stars (Tenenbaum et al.~\citeyear{tenenbaum}, De Beck et al.~\citeyear{debeck}, Ag\'undez et al.~\citeyear{agundez}), but never in dense star-forming cores. Due to this lack of observational constraints, the chemistry of P in the interstellar medium in general, and in star-forming regions in particular, has remained substantially unknown.

A considerable step forward was made in the last years thanks to the recent detection of PO in two high-mass (Rivilla et al.~\citeyear{rivilla2016})
and a low-mass (Lefloch et al.~\citeyear{lefloch}) star-forming regions. Moreover, with the IRAM-30m Telescope, Fontani et al.~(\citeyear{fontani2016}) detected PN (2--1) in 8 additional high-mass star-forming cores in different evolutionary stages: 2 starless cores (HMSC), 3 protostellar objects (HMPO), and 3 ultracompact HII regions (UCHII). 
One of the main findings of Fontani et al. (\citeyear{fontani2016}) is that all detected PN line widths are smaller than 5 km s$^{-1}$, and arise from regions associated with kinetic temperatures smaller than 100 K. Although the few detections reported in the literature are associated with warmer and turbulent sources, or even shocked material (Lefloch et al.~\citeyear{lefloch}), the results of Fontani et al. (\citeyear{fontani2016}) indicate that PN can arise also from relatively quiescent and cold gas. This information challenges theoretical models that invoke either high desorption temperatures or grain sputtering from shocks to release phosphorus into the gas phase (e.g. Turner et al.~\citeyear{turner}).
In this Letter, we present multi-transition observations of PN (2--1, 3--2, and 6--5) carried out with the IRAM 30m telescope towards our sample of massive dense cores. 
The main finding of this work firmly confirms our previous conclusion, i.e. that the origin of PN is not unique, because it may form in shocks but also in quiescent material.



\begin{figure}
\centering
\includegraphics[width=9cm,angle=-90]{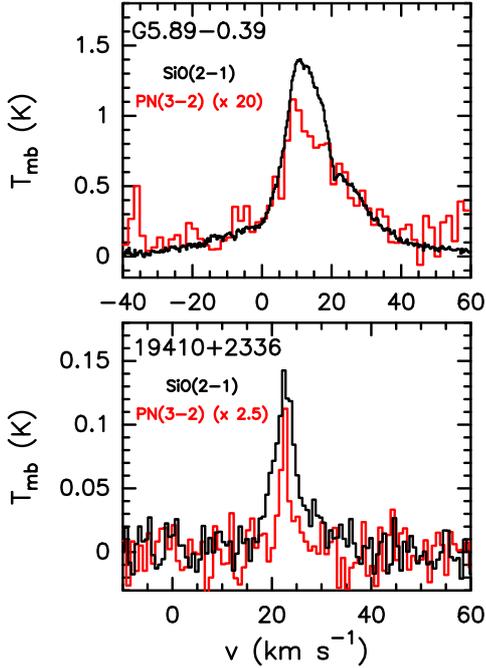}
 \caption{{\it Top panel:} PN (3--2) (red line; multiplied by a factor of 20) and SiO (2--1) (black line) lines measured toward the G5.89 UC HII region. {\it Bottom panel:} same as top panel for 19410+2336 (the PN emission is multiplied by a factor 2.5).}
    \label{fig-profiles}
\end{figure}



\begin{table}
	\begin{center}
	\caption{PN transitions observed in this work. We also include the SiO and SO transitions used in Sect.~\ref{discu}.}
        \tabcolsep 3.0pt
        \scriptsize
\begin{tabular}{c c c c c c} 
		\hline
		Molecule & Transition & Frecuency (GHz) & E$_{\rm up}$ (K) & A$_{ul}$ (s$^{-1}$) & $n_{\rm cr}$ (\cmc ) \\
		\hline
		PN & 2--1 & 93.97977  & 6.8  & 2.9$\times$10$^{-5}$ & (6.5-9.7)$\times 10^5$$^{(a)}$ \\
		PN & 3--2 & 140.96769 & 13.5 & 1.1$\times$10$^{-4}$ & (2.3-3.1)$\times 10^6$$^{(a)}$  \\
               PN & 6--5 & 281.91420   & 47.4 & 9.1$\times$10$^{-4}$ & (2.4-1.5)$\times 10^7$$^{(a)}$  \\
               SiO &  2--1 & 86.84696  & 6.3  &  2.9$\times$10$^{-5}$ & (3.6-5.9)$\times 10^5$$^{(b)}$ \\
               SiO &  5--4 & 217.10498  & 31.3  &  5.2$\times$10$^{-4}$ & (4.3-4.7)$\times 10^6$$^{(b)}$ \\
               SO & 2(2)--1(1) & 86.09395 & 19.3 & 5.3$\times$10$^{-6}$ & (1.8-3.8)$\times 10^6$$^{(b)}$  \\
		\hline
	\end{tabular}
\end{center}
\label{table-transitions}
$^{(a)}$ derived from the collisional rate coefficients in Tobola et al.~(\citeyear{tobola}) calculated in the range 10--80~K;\\
$^{(b)}$ derived from the collisional rate coefficients in the LAMBDA\footnote{http://home.strw.leidenuniv.nl/~moldata/} 
database calculated in the range 10--150~K;\\		
\end{table}
\normalsize

\begin{table*}
\caption{Source coordinates, and molecular parameters derived in this work. 
Col.~4 gives the line width at half maximum of PN (3--2) ($\Delta {\rm v}$); Cols.~5 -- 8 list the
integrated area of the PN (3--2) and (6--5) lines ($\int_{}T\ped{mb}\,dv$) calculated from gaussian fits
to the lines (when possible), the PN excitation temperatures ($T\ped{ex}$), and the PN total
column densities ($N\ped{\textit{tot}}$), respectively. Cols.~9 -- 12 give the
same parameters for SiO lines. $T\ped{ex}$ and $N\ped{\textit{tot}}$ are derived for
PN and SiO have been derived from the rotation diagram method assuming a source size of 9\asec.
Cols.~13 and 14 give integrated area of the SO 2(2)--1(1) line, and the SO total column 
density computed from it as explained in Sect.~\ref{analysis}. The errors on the integrated
areas are derived directly from the fit algorithm for gaussian lines, while for non gaussian lines
they have been calculated from the propagation of errors on the sum of the channels with 
signal (above 3$\sigma$ rms).}
\begin{center}
\scalebox{0.7}{
\renewcommand\arraystretch{1.3}
\begin{tabular}{lccccccccccccccc}
\hline \hline
 & & & \multicolumn{5}{c}{PN} & & \multicolumn{4}{c}{SiO} & & \multicolumn{2}{c}{SO} \\
 \cline{4-8}  \cline{10-13} \cline{15-16} \\
 & R.A.(J2000) & Dec.(J2000) & $\Delta {\rm v}$ (3--2) & $\int_{}T\ped{mb}\ap{3-2}\,dv$ & $\int_{}T\ped{mb}\ap{6-5}\,dv$& T\ped{ex} &N\ped{tot} & & $\int_{}T\ped{mb}\ap{2-1}\,dv$&  $\int_{}T\ped{mb}\ap{5-4}\,dv$& $T\ped{ex}$ & $N\ped{tot}$ & &$\int_{}T\ped{mb}\,dv$ &$N\ped{tot}$ \\ 
 & [h:m:s] & [${\circ}:{\prime}:{\prime \prime}$]  & \scriptsize{[km s$^{-1}$]} &\scriptsize{[K km/s]}&\scriptsize{[K km/s]}&\scriptsize{[K]} &\scriptsize{[10\ap{12}cm\ap{-2}]}& & \scriptsize{[K km/s]} &\scriptsize{[K km/s]} &\scriptsize{[K]} & \scriptsize{[10\ap{14}cm\ap{-2}]} & & \scriptsize{[K km/s]}& \scriptsize{[10\ap{14}cm\ap{-2}]}\\
            \hline
         \multicolumn{16}{c}{HMSC}\\
            \hline
 AFGL5142-EC  & 05:30:48.7 & +33:47:53 & 4.5$\pm$1.0  &0.20$\pm$0.03  &<0.24 &$3.2^{+0.8}_{-0.5}$&$9.0^{+5}_{-3}$	& &8.98$\pm$0.08\ap{ng} &14.18$\pm$0.09\ap{ng} &$8.0^{+0.6}_{-0.5}$&$1.6^{+0.2}_{-0.1}$& & 3.88$\pm$0.06\ap{ng} &    27.5$\pm$0.4 \\
 05358-mm3     & 05:39:12.5 &	+35:45:55 & 5.2$\pm$1.1  &0.14$\pm$0.02 &<0.16 &$4.9^{+2.7}_{-1.2}$&$2.2^{+1.0}_{-0.5}$  & &4.81$\pm$0.06\ap{ng} &3.32$\pm$0.06\ap{ng} &$6.3^{+0.4}_{-0.3}$&$0.86^{+0.11}_{-0.09}$ & & 1.68$\pm$0.05\ap{ng} &13.0$\pm$0.4 \\
  \hline
   \multicolumn{16}{c}{HMPO}\\
    \hline
AFGL5142-MM  & 05:30:48.0 & +33:47:54 & 2.9$\pm$0.8  &0.280$\pm$0.015\ap{ng} &<0.23 &$5.0^{+1.8}_{-1.0}$&$4.1^{+1.6}_{-0.9}$ & &12.24(0.08)\ap{ng} &13.62$\pm$0.10\ap{ng} &$7.2^{+0.4}_{-0.4}$&$2.2^{+0.2}_{-0.2}$ & & 4.73$\pm$0.05\ap{ng} &39.3$\pm$0.4 \\
  18089-1732     & 18:11:51.4 & $-$17:31:28 & 5.0$\pm$1.7   &0.37$\pm$0.03\ap{ng} &<0.22 & $32.1^{c}_{-22.5}$& 3.4\ap{d} & & 2.65$\pm$0.05\ap{ng} &5.96$\pm$0.07\ap{ng} &$9.0^{+0.8}_{-0.7}$&$0.50^{+0.05}_{-0.04}$ & & 2.23$\pm$0.04\ap{ng} &19.8$\pm$0.3 \\
  18517+0437$^{a)}$    & 18:54:14.2 &	+04:41:41 &1.9$\pm$0.3   &0.091$\pm$0.013 &0.063$\pm$0.013 & 10.4$\pm$1.2 & 0.7$\pm$0.2 & &2.10$\pm$0.03\ap{ng} &4.67$\pm$0.05\ap{ng} &$8.9^{+0.8}_{-0.6}$&$0.39^{+0.04}_{-0.03}$ & & 1.75$\pm$0.03 &  1.7$\pm$0.2\ap{e}  \\
    G75-core$^{b)}$       & 20:21:44.0 & +37:26:38 & 5.7$\pm$1.4  &0.088$\pm$0.018 &<0.14 & - & 0.60$\pm$0.12 & & 3.56$\pm$0.04\ap{ng}  &7.71$\pm$0.08\ap{ng} &$8.9^{+0.7}_{-0.7}$&$0.66^{+0.07}_{-0.06}$ & & 4.12$\pm$0.03\ap{ng} &	74.7$\pm$0.6 \\ 
\hline
 \multicolumn{16}{c}{UCHII}\\
    \hline
    G5.89-0.39$^{a)}$   & 18:00:30.5 & $-$24:04:01 & 17$\pm$3    &0.97$\pm$0.08\ap{ng} & 0.51$\pm$0.05 & 9.8$\pm$0.2 &7.4$\pm$2.0 & & 31.09$\pm$0.10\ap{ng}  &90.65$\pm$0.13\ap{ng}  &$9.9^{+0.9}_{-0.7}$&$6.0^{+0.5}_{-0.5}$ & & 15.30$\pm$0.08\ap{ng} & 121.6$\pm$0.6\ap{e}\\
 19410+2336    & 19:43:11.4 & +23:44:06 & 1.5$\pm$0.2    &0.080$\pm$0.010 &<0.15 & $4.6^{+2.0}_{-1.1}$ & $1.4^{+0.9}_{-0.4}$ && 0.76$\pm$0.04\ap{ng} & 1.18(0.04)\ap{ng}  & $7.9^{+0.8}_{-0.6}$&$0.139^{+0.02}_{-0.017}$ & & 1.53$\pm$0.03\ap{ng} &10.11$\pm$0.18 \\
  ON1$^{a)}$                & 20:10:09.1 & +31:31:36  &  5.5$\pm$0.8   &0.19$\pm$0.02 &0.12$\pm$0.03 & 9.1$\pm$0.8 &2.2$\pm$0.4 & &4.93$\pm$0.05\ap{ng}  &5.71$\pm$0.05\ap{ng}  &$7.2^{+0.5}_{-0.4}$&$0.89^{+0.10}_{-0.09}$ & & 1.84$\pm$0.03\ap{ng} & 13.3$\pm$0.2 \\
\hline \hline
\end{tabular}
\renewcommand\arraystretch{1.0}}
\end{center}
\justify
\ap{a} sources with PN (2-1), (3-2) and (6-5) rotational lines all detected;\newline
\ap{b} sources with only one PN line detected. It is not possible to derive $T\ped{ex}$. $N\ped{\textit{tot}}$ is calculated trough \mbox{Eq. (A4)} of Caselli et al.~(2002), assuming $T\ped{ex}$ equal to the mean of the temperature derived for the other sources in the same class;\newline
\ap{c} upper limit not defined, because the maximum slope line provides a non physical value of $T\ped{ex}$;\newline
\ap{d} lower limit not defined, because the maximum slope lines is not physically acceptable, while the upper limit value falls below the best fit, due to the difference in temperature between the minimum slope line and the best fit. For this source we calculated the total column density in a second way, using Eq. (A4) of Caselli et al.~(2002) with the value of the PN (3-2) area and with the temperature derived from the Boltzmann plot (T\ped{ex}=32.1 K). The result is $N\ped{tot}= 3.3\pm0.3$;\newline
\ap{e} $T\ped{kin}$ for NH\ped{3} is not available for these sources. The temperature used to derive the total column density is the mean of the $T\ped{kin}$, calculated separately in each evolutionary group;
\newline
\ap{ng} lines with non gaussian profile, for which the integrated area has been derived by summing the intensities of the channels with signal.\newline
\label{table-sample}
\end{table*}

\section{Observations}
\label{obs}

We have observed with the IRAM-30m Telescope the (3--2) and (6--5) rotational transitions of PN towards 12 sources of the Fontani et al. (\citeyear{fontani2011}) 
sample (the 8 objects detected in 2--1 in Fontani et al.~\citeyear{fontani2016} and four additional non-detections)
from the 6th to the 11th of December, 2016, using simultaneously the 2 and 0.8 mm 
bands of the EMIR receiver of the IRAM-30m telescope (IRAM-30m project n.~119-16).
Table~\ref{table-transitions} presents some main technical observational parameters, while the sources are listed in Table~\ref{table-sample}. 
The observations were made in wobbler-switching mode with a wobbler throw of 240\arcsec. 
The data were calibrated with the chopper wheel technique (see Kutner \& Ulich~\citeyear{kutner}), with a calibration uncertainty 
of about $10\%$. 
The spectra were obtained with the fast Fourier transform spectrometers with the finest spectral resolution (FTS50), 
providing a channel width of 50 kHz. All calibrated spectra were analysed using the GILDAS software developed 
at the IRAM and the Observatoire de Grenoble. The spectroscopic parameters used in the derivation of the column 
densities (Table~\ref{table-transitions}) have been taken from the Cologne Molecular Database for Spectroscopy (CDMS; M\"{u}ller et al. 2001, 2005).
Further details on the observations will be given in a forthcoming paper that presents an extended analysis of this
dataset (Mininni et al. in prep.). 




\section{Analysis}
\label{analysis}

The new observations show that PN(3--2) is detected in 9 sources: 8 previously detected in PN(2--1),
and a new one (G75-core).
The 3--2 transition is in all cases more intense than the 2--1, with a ratio of line intensities I$_{\rm 3-2}$/ I$_{\rm 2-1}$ 
between 1.5 and 3. The high-energy 6--5 transition is detected only towards 3 sources: 18517+0437, G5.89-0.39 and ON1.
Although the lines have hyperfine structure, the faintest components were either below the 3$\sigma$ level,
or blended among them, so that a fit simultaneous to all the hyperfine components could not be performed. 
Therefore, we have fitted the lines with single gaussians, and used the integrated areas obtained 
to construct the rotational diagrams. For non-gaussian profiles, the integrated areas have been derived by summing the 
intensities of the channels above the 3$\sigma$ rms level. The rotational diagrams have provided us with the 
PN excitation temperature, $T\ped{\rm ex}$, and the PN total column densities, $N\ped{\rm tot}$. 
The method assumes that the molecular levels are populated with a single excitation temperature. 
This is certainly a good approximation for the sources in which only the (2--1) and (3--2) lines, which have similar 
energy of the upper level, are detected. However, even in the three sources in which the (6--5) is included, 
the points in the rotational diagram are well distributed on a straight line, so that we are confident that the approximation
is reliable. The source angular size, needed to compare transitions observed with different beam sizes, is unknown so far.
Therefore, we have assumed that the PN emission in all the lines fills the smallest beam size, i.e.~9\asec . 
The integrated line intensities have then been corrected for beam dilution
to determine $N\ped{\rm tot}$. The $T\ped{\rm ex}$ and $N\ped{\rm tot}$ derived 
from the rotational diagrams, and the integrated areas used to make them, are shown in Table~\ref{table-sample}.
The spectra of all detected sources and the rotational diagrams are included in an Appendix available on-line only.
The measured $T\ped{\rm ex}$ are in the range $\sim 5$ to 30 K (see Table~\ref{table-sample}) which
is lower (in some cases much lower) than the gas kinetic temperatures given in Fontani et al.~(\citeyear{fontani2011}).
These latters are in the range $\sim 25 - 40$~K, and were derived from ammonia rotation temperatures.
which indicates that the PN lines are sub-thermally excited. This is consistent with the high 
critical density of the PN transitions ($\geq 10^{5-6}$~\cmc , Table~\ref{table-transitions}), 
which is higher than the average H$_2$ volume density expected at the angular resolution of 
our observations ($\sim 10^{4-5}$\cmc ). The method assumes also optically thin transitions. 
Given the low PN abundance, we are confident that this assumption is satisfied as well.
The derived total column densities are in the range $N\ped{\rm tot}$=$(0.6 - 9)\times 10^{12}$\cmq .
We have checked if our $N\ped{\rm tot}$ and those computed by Fontani et al.~(\citeyear{fontani2016}) are 
consistent. For this purpose, we have rescaled the values computed by Fontani et al.~(\citeyear{fontani2016}),
averaged within a beam of $\sim 26$\asec , to 9\asec . The two estimates are consistent within a factor 2.
The discrepancy might be due to the fact that the temperature we used in Fontani et al.~(\citeyear{fontani2016}),
i.e. the kinetic temperature obtained from ammonia, was too
high ($\sim 25 - 40$~K) for PN. Details on the calculation of the column densities source by source are also 
given in the caption of Table~\ref{table-sample}. 
We have also evaluated how our results could change in case of higher opacities of the
lines: we have taken the three sources in which we have detected
the three transitions, and assumed $\tau \sim 1$ for the (2--1) line: the total column densities 
can change by a factor 2, while the rotation temperatures of a factor 1.25. 
Therefore, even in the unfavourable, and unlikely, event of a higher opacity
of the lower excitation lines, the column densities should vary at most by a factor 2.
Finally, the PN line widths at half maximum $\Delta {\rm v}$ (Col.~4 in Table~\ref{table-sample}) are
generally around 5~\kms\ or larger, i.e. much larger than the thermal broadening (expected to
be of a few 0.1~\kms ) and thus associated with high-velocity or shocked gas, but not everywhere.
In fact, the (3--2) transitions of AFGL5142-MM, 18517+0437, and 19410+2336 have $\Delta {\rm v}\sim 2.9$,
$\sim 1.9$ and $\sim 1.5$ \kms , respectively. These values are lower than what is usually expected for shocked 
or very turbulent material.  We will discuss better this relevant point in the next Section.

\section{Discussion and conclusions}
\label{discu}

In Rivilla et al.~(\citeyear{rivilla2016}), based on the models of Vasyunin \& Herbst~(\citeyear{veh}) 
and on the detections of PN and PO in the two high-mass star-forming regions W51 and W3(OH), we
have concluded that the two molecules are formed during the cold pre--stellar phase, 
and then released in hot gas in the warm-up phase (after $\sim 35$~K), i.e. they should be a 
product of grain mantle evaporation. This interpretation was consistent with the broad lines
and the high gas turbulence found in these two objects, but it is not appropriate to explain the 
results of Fontani et al.~(\citeyear{fontani2016}), in which some PN lines are too narrow 
($\sim 1$~\kms ) to be produced in warm/hot gas. In order to better interpret the origin of PN, 
in this study we have analysed also the molecular species SiO and SO (shock tracers), which were 
observed simultaneously to the lines published by Colzi et al.~(\citeyear{colzi}) and Fontani et al.~(\citeyear{fontani2015a}). 
The spectroscopic parameters of the detected lines are listed in Table~\ref{table-transitions}. The analysis of 
the SiO transitions is similar to that performed for PN, while for SO, for which we have only one
line, the total column densities have been derived by assuming as excitation temperature the
kinetic gas temperature given in Fontani et al.~(\citeyear{fontani2011}). 
First of all, we have compared the line profiles, and found that in some sources SiO and PN are 
very similar, while in others the PN lines are clearly much narrower.
Fig. \ref{fig-profiles} show the comparison of the PN(3--2) and the SiO(2--1) line profiles towards the UCHII regions 
G5.89--0.39 and 19410+2336, which are the two objects in which this dichotomy is most apparent. 
The PN profile follows nicely the SiO profile in G5.89--0.39, in particular in the high velocity wings, suggesting a similar 
origin from shocked gas. This close relation between these two species was recently also seen in the 
protostellar shock L1157-B1 (Lefloch et al.~\citeyear{lefloch}). On the other hand, in 19410+2336, 
the PN profile is much narrower than that of SiO ($\Delta {\rm v}\sim $1.5~\kms\ against $\sim 5$~\kms), 
suggesting that at least in this source, and the others in which this difference is apparent, the 
PN emission must arise from more quiescent material. 
To better quantify this result, we have calculated the ratio between the $\Delta {\rm v}$
of SiO (2--1) and PN (3--2), and found that three sources,  AFGL5142-MM, 18517+0437, and 19410+2336, 
have $\Delta {\rm v}$(SiO)/$\Delta {\rm v}$(PN)$\sim 1.8$, 2.2, and 4.1, respectively. The others have
$\Delta {\rm v}$(SiO)/$\Delta {\rm v}$(PN)$\leq 1.5$. Based on this threshold, we distinguish between sources 
with "Narrow" (N) and "Broad" (B) PN lines. 
The three N objects are HMPOs or UCHIIs, and although high-velocity gas is present in the 
sources (as demonstrated by the broad SiO lines), PN certainly does not arise from this gas. 
We also note that the two HMSCs detected in PN are both B sources, despite the lower turbulence 
and the absence of shocked material
expected in starless cores. However, we know that the outer envelope of both sources is perturbed by 
external nearby sources (see Fontani et al.~\citeyear{fontani2011} for details), and our data encompasses 
a region with an angular size that includes this envelope. Therefore, we think that the broad PN
emission likely arises from the perturbed envelope, but only higher angular resolution 
observations can conclusively confirm this.

\begin{figure}
\centering
{\includegraphics[width=4.2cm,angle=-90]{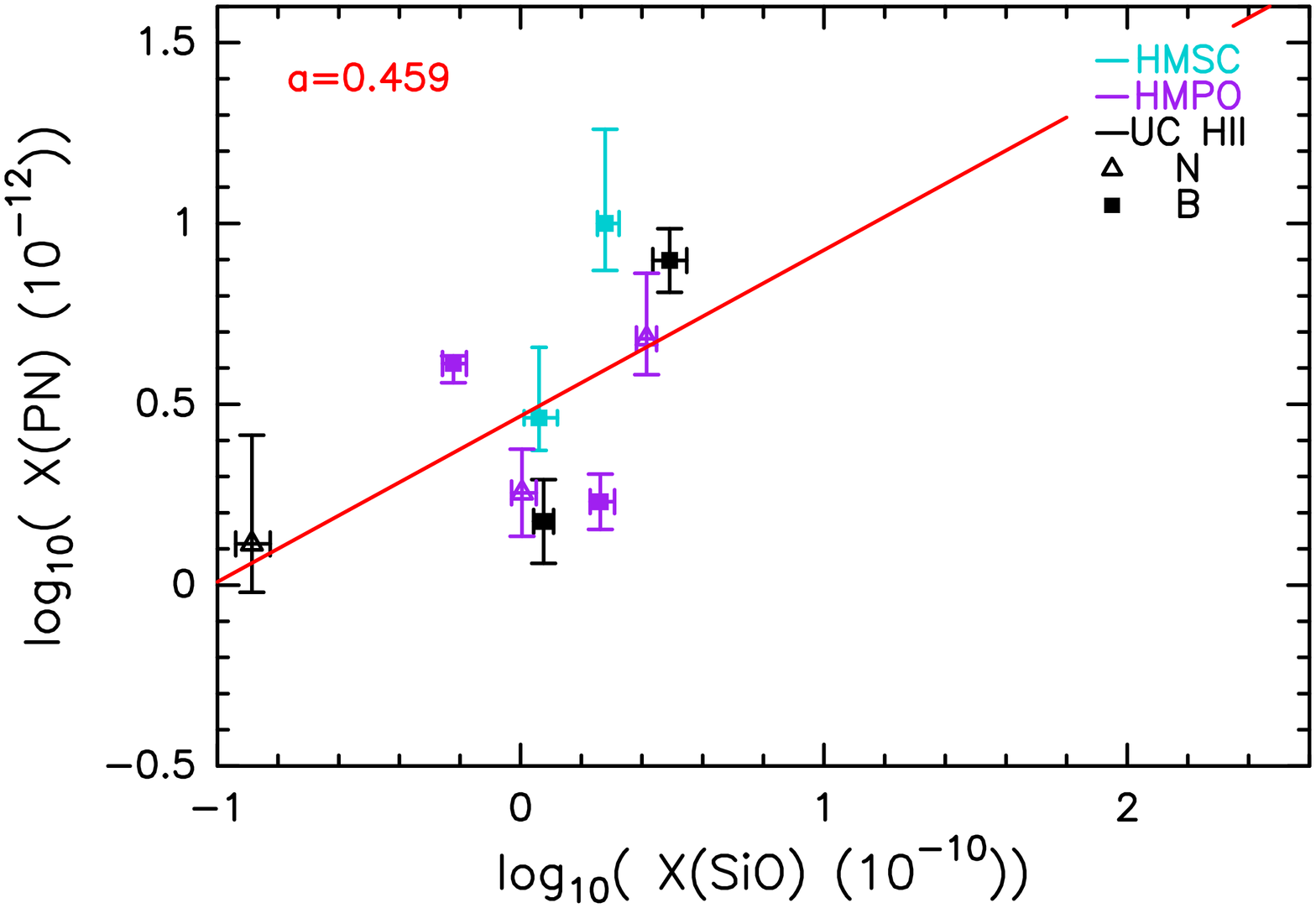}}
{\includegraphics[width=4.2cm,angle=-90]{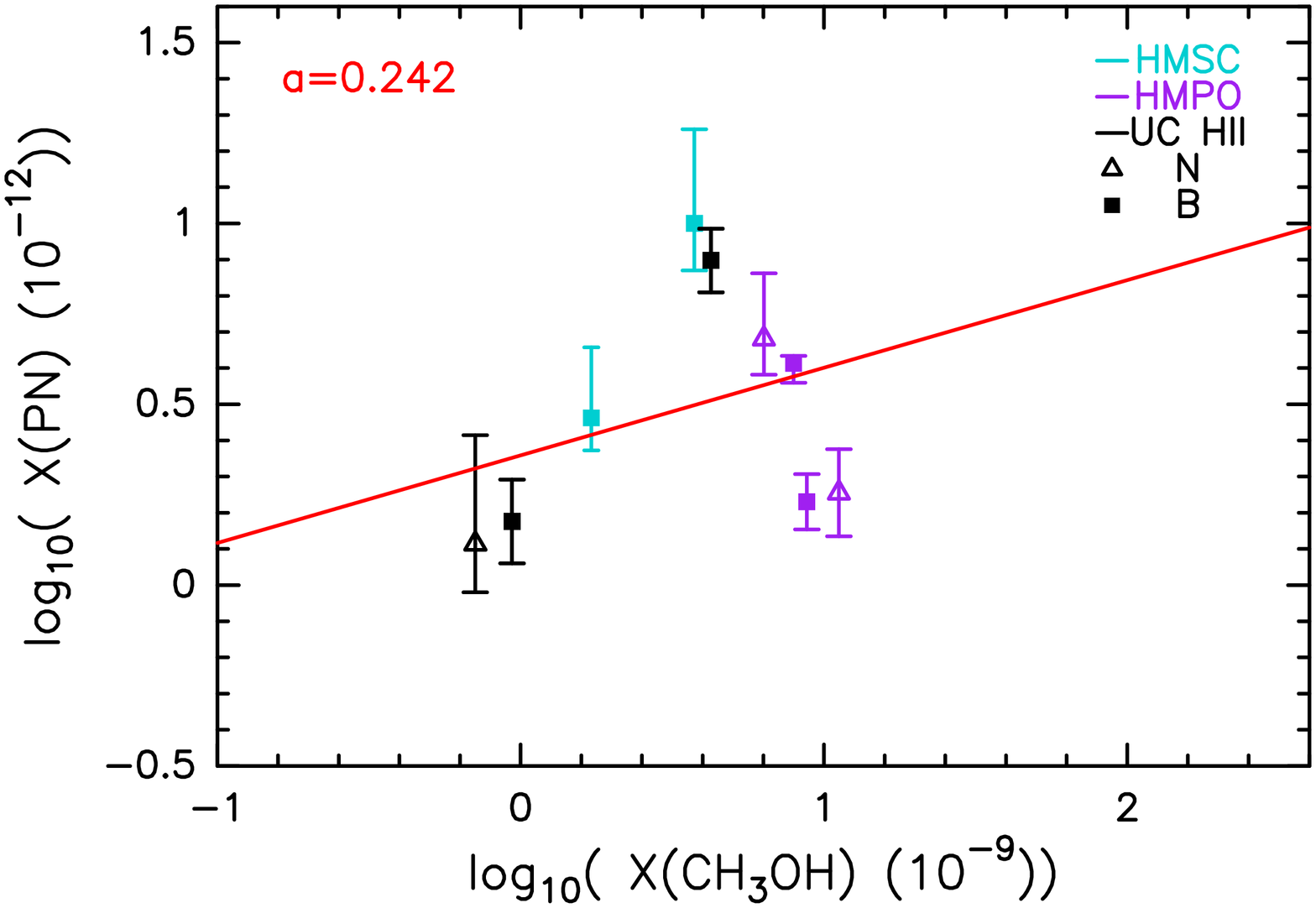}}
{\includegraphics[width=4.2cm,angle=-90]{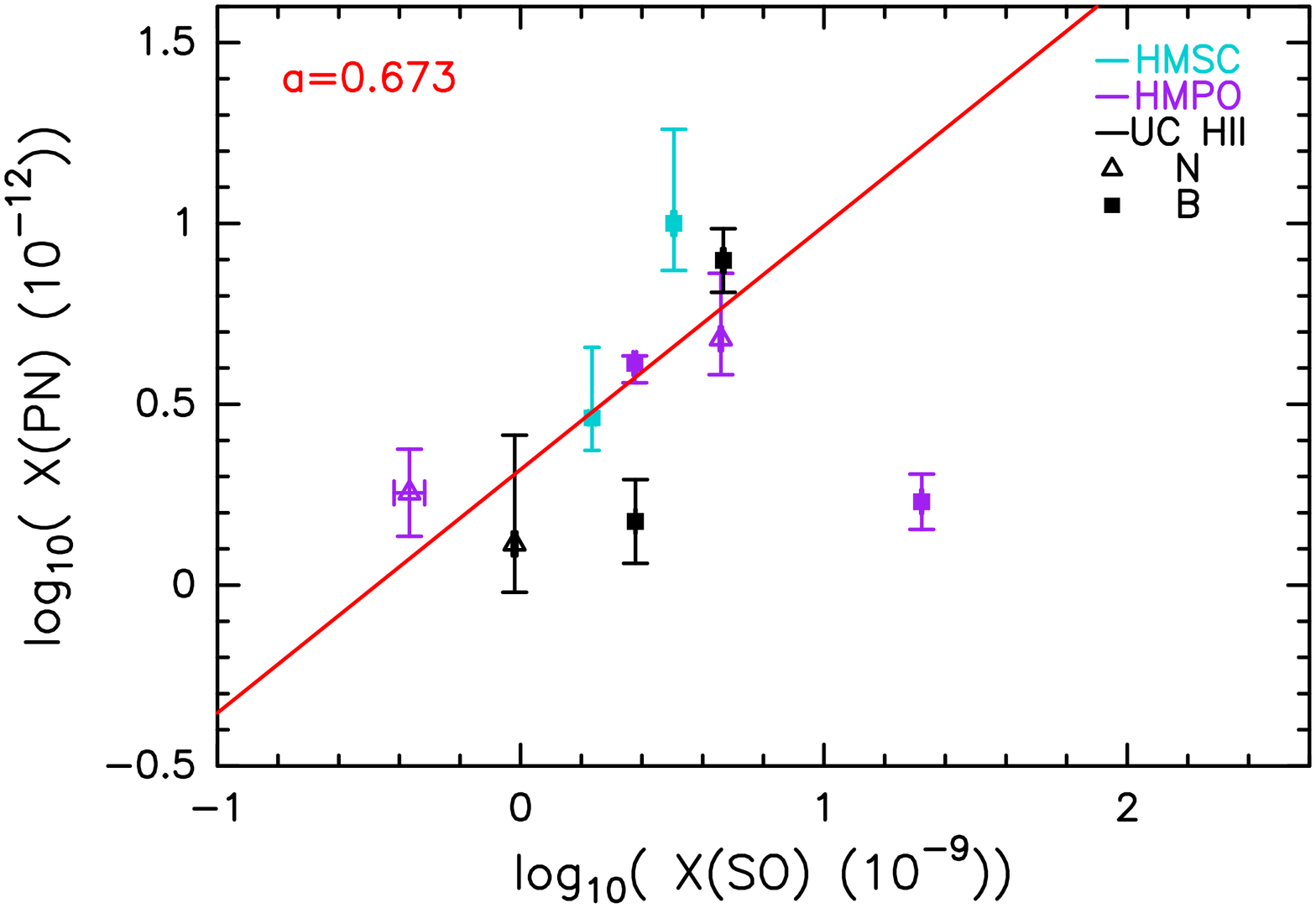}}
{\includegraphics[width=4.2cm,angle=-90]{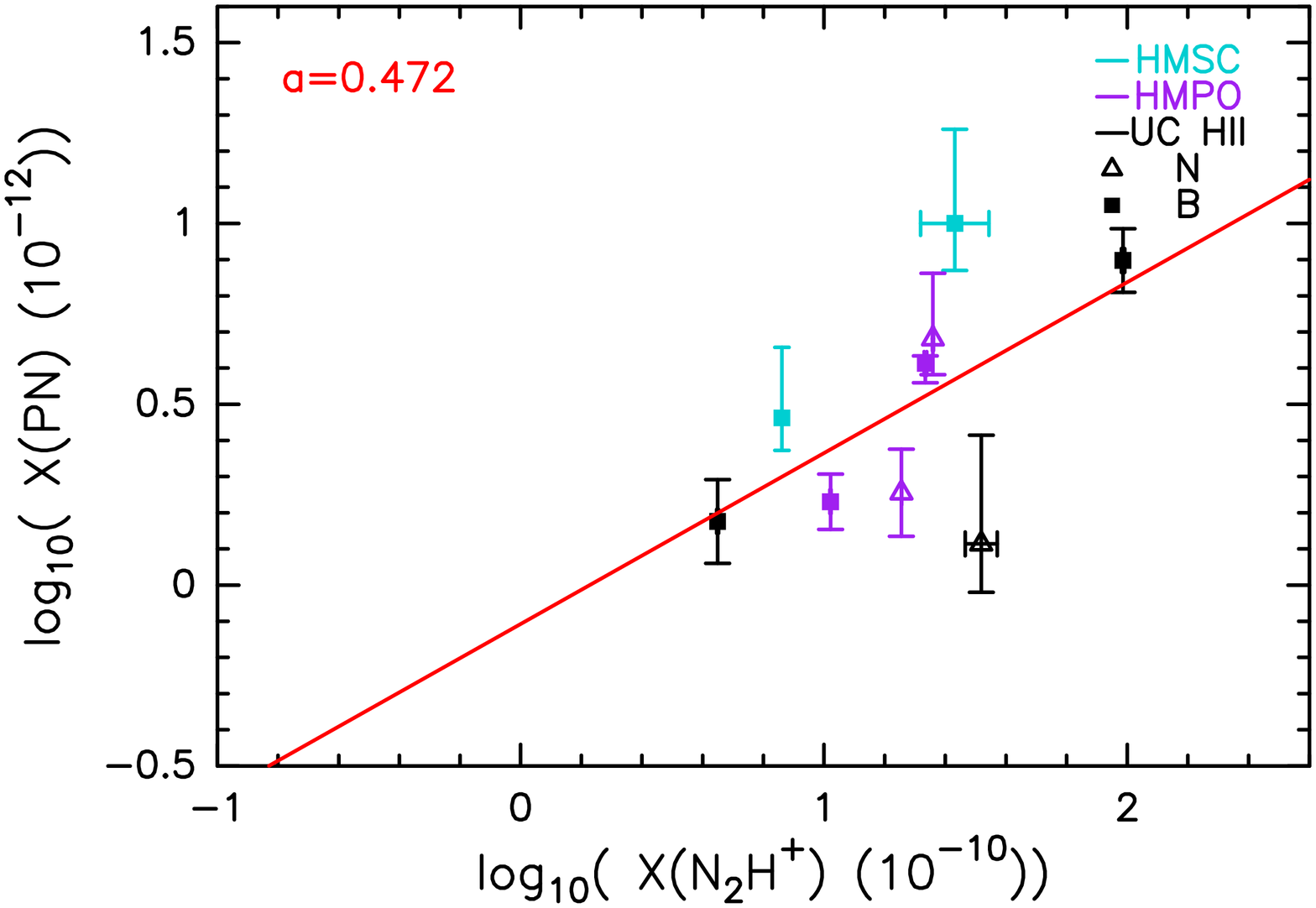}
}
 \caption{Abundance of PN calculated as explained in Sect.~\ref{appA} against the
 abundances of (from top to bottom): SiO, \METH , \H , and SO. The 
 squares indicate the "Broad" sources, and the triangles the "Narrow" sources. 
 In the panel with SO data, the G75-core, identified as an outlier
 by applying a Kolmogorov-Smirnov test, has been excluded from the linear fit. 
 The different colours indicate the three evolutionary stages as labelled in the
 top right corner of each frame. The slope in each plot is indicated in the top-left corner.
 }
    \label{fig-correlations}
\end{figure}

We have also derived the molecular abundance of PN relative to H$_2$, and compared it with that of 
SiO and SO (shock tracers), CH$_3$OH (product of surface chemistry, enhanced presence in HMPO and 
UCHII regions), and \H\ (produced by gas-phase chemistry). The \METH\ total column densities 
have been taken from Fontani et al.~(\citeyear{fontani2015a}), and those of \H\ from Fontani et 
al.~(\citeyear{fontani2015b}, from the 1--0 line). The derivation of the H$_2$ column densities, used to
compute the molecular abundances, is described in Appendix~\ref{appA}. The comparison among the
different abundances is shown in Fig.~\ref{fig-correlations}. In order to estimate a possible relation 
between the various species, we have performed a linear fit to the data. Although the statistics 
is low, we have found a positive trend
with SiO, SO (only if we exclude from the statistical analysis the outlier G75-core), and \H , 
while with \METH\ a correlation is tentative at best. The slope of the linear fits are given
in Fig.~\ref{fig-correlations}. 
The trend does not improve considering only the B sources, as can be noted from 
Fig.~\ref{fig-correlations}. In summary, the similar SiO and PN line profiles in six targets 
(see Fig.~\ref{fig-profiles}), corroborated by a faint but positive trend between their abundances, 
indicate that a source of PN is certainly in shocked gas. However, the different (narrower) PN 
profiles found in three objects where shocks are at work rule out the possibility that this is the 
unique origin, and indicate alternative formation routes in more quiescent material. These conclusions 
need to be supported by a better statistics and, above all, higher angular resolution observations, 
without which the nature of the PN emission cannot be conclusively established.

\vspace{0.2cm}
\noindent
We have presented multi-transitions observations of PN (2--1, 3--2 and 6--5) carried out with the IRAM 30m telescope 
towards our sample of massive dense cores in different evolutionary stages. We found that the excitation 
temperatures of PN are low ($\sim $5--30 K) even in HMPOs and UCHIIs. 
This suggests that PN is sub-thermally excited because the average density of the regions (typically 
10$^{4-5}$ cm$^{-2}$) is lower than the critical density of PN ($>$ 10$^{5-6}$ cm$^{-2}$). 
We have found a good agreement between the line profiles of PN and those of the well known 
shock-tracer SiO in six sources, which suggests that in these objects PN is likely sputtered from dust 
grain mantles through shock waves. This conclusion is in good agreement with the recent results 
found in a survey of Galactic centre clouds (Rivilla et al~\citeyear{rivilla2018}). However, in 
three objects the PN lines are at least 1.8 times narrower than those of SiO, suggesting that the 
emission must arise from more quiescent gas.
Our study, based on single-dish observations, will be implemented by on-going higher-angular 
resolution observations that will reveal the spatial distribution of PN, necessary to support
our conclusions and evaluate the "fraction" of PN produced in the two ways.

\section*{Acknowledgements}

We thank the anonymous referee for his/her constructive comments. 
We thank the IRAM-30m staff for the support during the observations.
V.M.R. acknowledges the financial support received from the European Union's Horizon 2020 research 
and innovation programme under the Marie Sklodowska-Curie grant agreement No 664931, and from 
the Italian  Ministero  dell'Istruzione, Universit\`a e Ricerca through the grant Progetti Premiali 
2012 - iALMA (CUP C52I13000140001). P.C. and A.V. acknowledge support from the European 
Research Council (ERC project PALs 320620) 








\appendix

\section{Derivation of H$_2$ column densities and molecular abundances}
\label{appA}

\begin{table}
	\begin{center}
	 \scriptsize
	\caption{Sample: continuum flux densities, $S_{\nu}$, derived hydrogen column densities, 
	$N$(H$_2$), and parameters used to compute it (i.e.~distance and temperature, from Fontani et al.~\citeyear{fontani2011}).
	$N$(H$_2$) is a beam-averaged value (14\asec\ for SCUBA, 19.2\asec\ for ATLASGAL, 11\asec\ for CSO).}
       \tabcolsep 10.0pt
	\begin{tabular}{c c c c c}
    \hline
		Source & $S_{\nu}$ & $N$(H$_2$) & $d$ & $T$ \\ 
 		       & [Jy]   & [10$^{23}$ cm$^{-2}$]  & [kpc] & [K] \\   
		       \hline 
\multicolumn{5}{c}{HMSC}   \\      
		AFGL5142-EC     & 4.0(0.4) & 3.55 & 2.14$^{(a)}$ & 25 \\    
    	05358-mm3       &  4.5(0.2)  & 3.13 & 1.8 & 30 \\  
\hline
\multicolumn{5}{c}{HMPO}   \\                
        AFGL5142-MM     & 6.0(0.2)   & 3.55  & 2.14$^{(a)}$ & 34 \\   
        18089--1732     & 6.7(0.3) &  3.44 & 3.6 & 38 \\  
        18517+0437     & 3.4(0.2)$^{(b)}$ & 0.85$^{(b)}$ & 2.9 & 44 \\  
        G75-core       &  8.3(0.2)   & 1.47  & 3.8 & 96 \\
\hline
\multicolumn{5}{c}{UCHII}     \\          
        G5.89-0.39     &  17.3(0.7) &  21.1  & 1.3 & 20 \\
        19410+2336     &  3.85(0.05)  & 4.38  & 2.1 & 21 \\
        ON1           &  28$^{(c)}$  & 23.6$^{(c)}$  & 2.5 & 26 \\
\hline
	\end{tabular}
	\end{center}
\label{table-hydrogen}
$^{(a)}$ from Burns et al.~(\citeyear{burns}); \\
$^{(b)}$ from APEX (ATLASGAL) data at 870 $\mu$m or 345~GHz (http://www3.mpifr-bonn.mpg.de/div/atlasgal/index.html); \\
$^{(c)}$ from CSO data at 350 $\mu$m (Hunter et al.~\citeyear{hunter}).\\
\end{table}
\normalsize


We have computed the H$_2$ total column densities of the sources of our sample 
from the maps of Di Francesco et al. (2008),
who have measured the dust emission at $\sim 850$ $\mu$m (or $\sim 353$~GHz) with 
SCUBA at the JCMT, with an angular resolution of $\sim 14$\asec . For 18517+0437 and
ON1, absent in Di Francesco et al. (2008), we have used the images at $875$ $\mu$m
of the APEX ATLASGAL survey (http://www3.mpifr-bonn.mpg.de/div/atlasgal/index.html), 
and those otained with the CSO at $350$ $\mu$m by Hunter et al.~(\citeyear{hunter}). 
N(H$_2$) was derived at all wavelengths from the equation:
$M_{\rm dust} = \frac{S_{\nu}d^2}{\kappa_{\nu}B_{\nu}(T)}$
where $S_{\nu}$ is the total integrated flux density at frequency $\nu$, 
$d$ is the source distance, $\kappa_{\nu}$ is the dust mass opacity coefficient,
extrapolated from the value of 1 cm$^2$ gr$^{-1}$
at 250~GHz (Ossenkopf \& Henning~\citeyear{oeh}, assuming a
dust opacity index $\beta = 2$, i.e. a spectral index $2+\beta=4$), 
and $B_{\nu}(T)$ is the Planck function at dust temperature $T$. The
equation is valid for optically thin dust emission. Then, we have computed
the H$_2$ mass by multiplying $M_{\rm dust}$ for a mass gas-to-dust
ratio of 100, from which we have computed the average H$_2$ volume
density assuming a spherical source, and finally computed
$N({\rm H_2})$ by multiplying the volume density for the average
diameter of the sources (assumed to be equal to the beam size). 
We used as dust temperatures the kinetic temperatures
given in Fontani et al.~(\citeyear{fontani2011}). The derived values of the hydrogen 
column densities, averaged within the beam sizes, are shown in Table \ref{table-hydrogen}.
Finally, the molecular abundances shown in Fig.~\ref{fig-correlations} have been
then computed from the ratio between the molecular $N\ped{tot}$ and $N$(H$_2$),
after rescaling $N\ped{tot}$ to 14\asec\ (19.2\asec\ for 18517+0437, 11\asec\ for ON1).

\bsp	
\label{lastpage}
\end{document}